\documentclass[12pt]{article}
\usepackage{mathrsfs}
\usepackage{natbib,graphicx,setspace,lscape,longtable,epstopdf,xr}
\usepackage{mathrsfs,amsmath,amsthm,amssymb,color}
\usepackage{natbib,epsfig,graphicx,pdfpages}
\usepackage{rotating}
\usepackage{float}
\usepackage{bm}
\usepackage{CJK}
\usepackage{ragged2e}
\usepackage{threeparttable}
\usepackage{multirow}
\usepackage{url}
\usepackage{natbib,graphicx,setspace,lscape,longtable,epstopdf,xr,rotating,epsfig,pdfpages,float,bm,ragged2e,geometry, subfigure}
\usepackage{mathrsfs,amsmath,amsthm,amssymb,color}
\usepackage{threeparttable,booktabs,multirow}
\usepackage[font=small,labelfont=bf]{caption}
\usepackage{algorithm}
\usepackage{algorithmic}

\def\bt{\bm{\theta}}
\def\bl{\bm{\lambda}}
\def\bPsi{\bm{\Psi}}
\def\bPhi{\bm{\Phi}}

\bibpunct{(}{)}{;}{a}{,}{,}

\newtheorem{theorem}{Theorem}
\newtheorem{lemma}{Lemma}
\newtheorem{proposition}{Proposition}

\newcommand{\csection}[1]
    {\begin{center}
        \stepcounter{section}
        {\bf\large\arabic{section}. #1}
    \end{center}
}

\newcommand{\scsection}[1]
    {\begin{center}
        {\bf\large #1}
    \end{center}
}

\newcommand{\csubsection}[1]{
\begin{center}
\stepcounter{subsection}
{\it\arabic{section}.\arabic{subsection}. #1}
\end{center}
}

\def\beq{\begin{equation}}
\def\eeq{\end{equation}}
\def\beqr{\begin{eqnarray}}
\def\eeqr{\end{eqnarray}}
\def\beqrs{\begin{eqnarray*}}
\def\eeqrs{\end{eqnarray*}}
\def\bet{\begin{theorem}}
\def\eet{\end{theorem}}
\def\bel{\begin{lemma}}
\def\eel{\end{lemma}}
\def\bep{\begin{proposition}}
\def\eep{\end{proposition}}
\def\bg{\begin{figure}[tbph]\begin{center}}
\def\eg{\end{center}\end{figure}}

\def\bc{\begin{center}}
\def\ec{\end{center}}

\def\mL{\mathcal L}

\newcommand{\btt}[1]{\bt^{(#1)}}
\newcommand{\blt}[1]{\bl^{(#1)}}
\newcommand{\betat}[1]{\bm\beta^{(#1)}}
\newcommand{\parfrac}[2]{\frac{\partial {#1}}{\partial {#2}}}
\def\d{\mathrm{d}}

\numberwithin{equation}{section}

\usepackage[dvipsnames]{xcolor}

\begin{document}

\begin{center}
    {\bf\Large Implicit Profiling Estimation for Semiparametric Models with Bundled Parameters}

    Yucong Lin$^{1}$, Jinhua Su$^{2}$, Yang Liu$^{2}$, Jue Hou$^{3}$, Feifei Wang$^{2}$

{\it\small
$^1$ Beijing Engineering Research Center of Mixed Reality and Advanced Display, \\ School of Optics and Photonics, Beijing Institute of Technology, Beijing, China;\\
$^2$ Center for Applied Statistics and School of Statistics, \\ Renmin University of China, Beijing, China;\\
$^3$ Harvard T.H. Chan School of Public Health,
Department of Biostatistics.
}

\end{center}

\begin{singlespace}
    \begin{abstract}
Solving semiparametric models can be computationally challenging
because the dimension of parameter space may grow large with
increasing sample size.
Classical Newton's method becomes quite slow and unstable with intensive calculation of the large Hessian matrix
and its inverse.
Iterative methods separately update parameters for finite dimensional component and infinite dimensional component
have been developed to speed up single iteration, but
they often take more steps until convergence or even sometimes sacrifice estimation precision
due to sub-optimal update direction.
We propose a computationally efficient implicit profiling algorithm that achieves simultaneously the fast iteration step in iterative methods and the optimal update direction in the Newton's method by profiling out the infinite dimensional component
as the function of the finite dimensional component.
We devise a first order approximation when the profiling function has no explicit analytical form.
We show that our implicit profiling method always solve any local quadratic programming problem in two steps.
In two numerical experiments under semiparametric transformation models and GARCH-M models, we demonstrated
the computational efficiency and statistical precision of
our implicit profiling method.
    \end{abstract}


\end{singlespace}

\newpage

\csection{INTRODUCTION}

Semiparametric models are attractive choices for robust statistical analysis, as they identify the parameters of interest with minimal model assumption.
Semiparametric models have one parametric component
for parameter of interest
and another nonparametric component to promote
the flexibility of the model \citep{2012Asymptotic}.
The price for such flexibility includes the computational burden of solving the model.
The estimation of semiparametric models often involves the finite dimensional approximation of nonparametric component, whose dimension in turn grows with
the sample size.
As the number of parameter grows large,
the estimation problem becomes increasingly computationally challenging.

We focus on the ``\emph{bundled together}" cases \citep{huangIntervalCensoredSurvival1997}
in which the estimation of the parametric component and nonparametric component cannot be clearly separated.
In this work, we classify the bundled relationship between parametric and nonparametric components into two different types, i.e., the \emph{explicitly bundled} type and \emph{implicitly bundled} type. Specifically, the explicitly bundled type refers to the situation where the nonparametric component can be profiled out by an explicit expression of the parametric component. One typical example of this type occurs for missing data and censoring data. To estimate the probability of data missing or censoring, an unknown function consisting of the finite-dimensional parameters along with some covariates are often applied \citep{huangIntervalCensoredSurvival1997,Ding2011A,Chen2015Efficient}. In time series researches, the unknown relationship between conditional variance and mean is often regarded as the nonparametric component \citep{Christensen2012}. Other examples include, the single index model and the Cox regression model with unspecified link functions \citep{chen2002double,breslow2007weighted}, and the models with generated covariates \citep{tao1999estimation,2016Semiparametric, Frazier2018A}.

To estimate the semiparametric model with explicitly bundled parameters, we can plug in the explicit expression of the nonparametric component into the objective function (i.e., the loss function or the likelihood function)
or the estimating equations. As a result, the semiparametric model is converted into a parametric model. However, there exist situations that the explicit expression is not smooth or continuous \citep{chen2003estimation}. In this case, the numerical computational methods, such as the Newton-Raphson method, might suffer from the local optimal problem. In other words, the derivative of the objective function may crash in the neighbourhood of the jump point \citep{shi2020two}. To address this problem, several approximation solutions have been proposed, all of which target to find implicit substitutions for the initial derivative \citep{fan2007maximization,benaglia2009like}.

The implicitly bundled type refers to the case that the nonparametric component cannot be profiled out by an explicit expression of the parametric component. Then, the plug-in methods cannot be applied for estimation of semiparametric models in this case \citep{chen2017semiparametric}. To address this issue, various recursive updating methods have been proposed \citep{nawata1994estimation,terzija2003improved,jiang2020recursive}. The basic idea for recursive methods is to update the parametric component and nonparametric component iteratively in turn \citep{mentre1995two}. The convergence of the recursive methods have been theoretically guaranteed. However, to update each component separately, the recursive methods have to ignore the interaction between the two components. As a consequence, the combination of two separate update directions can deviate from the optimal update directions when the interaction is strong. As the result, the recursive updating algorithms may take more steps to converge.

In this work, we focus on the investigation of statistically and computationally efficient estimation method for the semiparametric models with both explicitly and implicitly bundled parameters. Motivated by the plug-in method commonly used in the explicit bundled type, we propose a novel implicit profiling method. Specifically, regardless of the types of bundled parameters, we seek to profile out the nonparametric component as a function of the parametric component.
The functions characterizing their relationship can be extracted from the first-order derivative of the objective function with respect to the nonparametric component. Next, plugging in the new functions into the first-order (i.e., the gradient) and second-order (i.e., the Hessian matrix) derivatives of the objective function with respect to the parametric component. The resulting new gradient and Hessian matrix of the parametric component then account for the interaction with nonparametric component. Based on the new gradient and Hessian matrix, the recursive updating strategy can be applied for model estimation.

To summarize, we provide the following contributions in this work. First, we classify semiparametric models with bundled parameters into the explicitly bundled type and implicitly bundled type. For model estimation, we develop an implicit profiling method, which can solve the estimation problem of both types. Second, in the implicit profiling method, we regard the nonparametric component as functions of the parametric component, which can be extracted from the gradients of the objective function. Then, we compute new implicit profiling gradient and Hessian matrix for the parameters of interest. We show theoretically that, the updating directions using implicit profiling gradient and Hessian matrix are in accordance with those computed by the Newton-Raphson method, which guarantees the estimation efficiency of the parameters of interest. Third, the implicit profiling method also behaves computationally efficient, when compared with the Newton-Raphson method. This is because, it applies the recursive updating strategy and thus avoids calculating the full Hessian matrix for both the finite-dimensional parameters and infinite-dimensional functions. Finally, we provide both explicitly bundled and implicitly bundled examples to demonstrate the estimation performance of the proposed implicit profiling method. Compared with the recursive iteration method and some state-of-the-art methods, the implicit profiling method still shows advantages in both estimation efficiency and computational efficiency.

The rest of this paper is organized as follows. Section 2 introduces the implicit profiling method in details. Section 3 illustrates the idea and algorithm of the implicit profiling method by a toy example. Section 4 takes the semiparametric transformation model as the example of the implicitly bundled type, shows application of the implicit profiling method under this model and compares its estimation performance using finite-sample simulations. Section 5 provides the semiparametric GARCH-M model as the example of explicitly bundled type. The estimation performance of the implicit profiling method is also verified through numerical studies. Section 6 concludes the paper with a brief discussion.

\csection{THE IMPLICIT PROFILING METHOD}

\csubsection{Preliminaries}

Assume we have a convex objective function  $\mathcal{L}(\bt, \bl)$, where $\bt$ is the parametric component with fixed dimension and $\bl$ is the parameter for finite-dimensional approximation of nonparametric component whose dimension may grow with the sample size.
Further assume $\bt$ and $\bl$ are bundled in this objective function, i.e., $\bt$ and $\bl$ cannot be clearly separated. Typical examples of this case include the semiparametric transformation model \citep{chanRiskPredictionImperfect2021} and the semiparametric GARCH-in-mean model \citep{Christensen2012}, which are discussed in details in Section 4 and 5. It is worth noting that, $\bl$ can also be a smooth function with infinite dimension. Then one can apply semiparametric methods to estimate $\bl$.

To estimate $\bt$ and $\bl$, let $\bPsi(\bt, \bl)$ and $\bPhi(\bt, \bl)$ denote the estimation equations with respect $\bt$ and $\bl$, respectively. Specifically, $\bPsi(\bt, \bl)$ is the derivative of $\mL(\bt, \bl)$ with respect to $\bt$. When $\bl$ is the nuisance parameter, then $\bPhi(\bt, \bl)$ is the derivative of $\mL(\bt, \bl)$ with respect to $\bl$ . In the case that $\bl$ is the smooth function, then $\bPhi(\bt, \bl)$ denotes the semiparametric estimation formula, such as the kernel smooth method or spline function. Then, to estimate $\bt$ and $\bl$, we need to solve:
\begin{equation}
\label{eq:object}
    \left\{
    \begin{aligned}
        &\bPsi(\bt,\bl) = \mathbf{0}\\
        &\bPhi(\bt, \bl) = \mathbf{0}.\nonumber
    \end{aligned}
    \right.
\end{equation}

\noindent
Let $\mathbf{G}(\bt, \bl) = (\bPsi(\bt, \bl)^{\top}, \bPhi(\bt, \bl)^{\top})^{\top}$. The entire updating algorithm (e.g, the Newton-Raphson method) requires $\mathbf{G}(\bt, \bl)=\mathbf{0}$. Let $\bm\beta = (\bt^{\top}, \bl^{\top})^{\top}$. Then the updating formula is $\bm\beta^{(k+1)} = \bm\beta^{(k)} - (\partial \mathbf{G}(\bm\beta^{(k)})/\partial \bm\beta)^{-1}\mathbf{G}(\bm\beta^{(k)}) $.
However, due to the bundled relationship of $\bt$ and $\bl$, the two parameters are often difficult to separate. This would result in a dense Hessian matrix $\partial \mathbf{G}/\partial \bm\beta$. Consequently, the calculation of the inverse of the Hessian matrix often suffers from high computational cost, which makes the entire updating algorithm computationally very inefficient.

To solve this problem, many recursive updating methods have been applied \citep{nawata1994estimation,terzija2003improved,jiang2020recursive}. Basically, the recursive method breaks the connection of $\bt$ and $\bl$ in the Hessian matrix. In other words, it considers the second-order derivative of the objective function with respect to each parameter, separately. Specifically, the updating formulas for the recursive method is given below:
\begin{equation}
\label{eq: simple iteration}
    \left\{
    \begin{aligned}
        &\bl^{(k+1)} = \bl^{(k)} - \left(\frac{\partial \bPhi(\bt^{(k)}, \bl^{(k)})}{ \partial \bl}\right)^{-1}\bPhi(\bt^{(k)}, \bl^{(k)})\\
        &\bt^{(k+1)} = \bt^{(k)} - \left(\frac{\partial \bPsi(\bt^{(k)}, \bl^{(k+1)})}{ \partial \bt}\right)^{-1}\bPsi(\bt^{(k)}, \bl^{(k+1)}).
    \end{aligned}
    \right.
\end{equation}
Although the update for $\bl$ is still a sub-problem with growing dimension,
the sub-problem Hessian $\partial \bPhi/\partial \bl$ is often sparse
by the design of the finite dimensional approximation of the nonparametric component --- different element in $\bl$ usually corresponds to the value of the nonparametric component at different locations.
Consequently, inverting $\partial \bPhi/\partial \bl$ can be much faster than inverting $\partial \mathbf{G}/\partial \bm\beta$.
The updating formula in \eqref{eq: simple iteration} iterates each parameter without considering the interaction with the other parameter. In other words, it makes approximation to the true second-order derivative $\partial \mathbf{G}/\partial \bm\beta$ by setting the off-diagonal blocks zero. For example, when updating $\bt$, the derivative $\partial \bPsi/\partial \bt$ considers $\bl$ as a constant and does not take into account the current value of $\bl$. Under the situation that $\bt$ and $\bl$ are strongly correlated with each other, the recursive method would definitely loss information and thus result in sub-optimal update directions.

\csubsection{Implicit Profiling Algorithm}

Both the entire updating method and recursive method are computationally inefficient. To address this issue, we propose an implicit profiling method. It is notable that, $\bt$ is the only parameter of interest. Therefore, we only focus on the efficient estimation of $\bt$. Recall that, in the recursive method, the updating formula for $\bt$ has lost some information by treating $\bl$ as a constant, which makes the estimation of $\bt$ inefficient. To address this problem, we propose an implicit profiling (IP) method. Specifically, we regard $\bl$ as the function of $\bt$, which we denote by $\bl(\bt)$. Then, the second-order derivative of the objective function respect to $\bt$ can be derived as follows:
\begin{equation}
\label{eq:ip_update}
    \frac{\partial \bPsi(\bt, \bl(\bt))}{\partial \bt} = \frac{\partial \bPsi(\bt, \bl(\bt))}{\partial \bt} + \frac{\partial \bPsi(\bt, \bl(\bt))}{\partial\bl(\bt)}\frac{\partial\bl(\bt)}{\partial\bt},
\end{equation}
where the derivative relationship $\partial\bl/\partial\bt$ can be obtained by solving $\partial \bPhi(\bt, \bl(\bt))/\partial\bl = \mathbf{0}$. We refer to \eqref{eq:ip_update} as the implicit profiling Hessian matrix of $\bt$, which decides the updating direction of $\bt$ in the implicit profiling method. Based on \eqref{eq:ip_update}, we can update $\bt$ and $\bl$ iteratively using the following updating formulas:
\begin{equation}
\begin{split}
\label{eq:ip_update2}
\bl^{(k+1)} &= \bl^{(k)} - \left(\frac{\partial \bPhi(\bt^{(k)}, \bl^{(k)})}{\partial \bl}\right)^{-1}\bPhi(\bt^{(k)}, \bl^{(k)})\\
\bt^{(k+1)} &= \bt^{(k)} -\left( \frac{\partial \bPsi(\bt^{(k)}, \bl^{(k+1)})}{\partial \bt} + \frac{\partial \bPsi(\bt^{(k)}, \bl^{(k+1)})}{\partial \bl}\frac{\partial\bl^{(k+1)}}{\partial\bt}\right)^{-1}\bPsi(\bt^{(k)}, \bl^{(k+1)}).
\end{split}
\end{equation}
\noindent
The complete algorithm of the implicit profiling method is present in Algorithm \ref{ag:generalization IP}.
\begin{algorithm}[h]
\setstretch{1}
    \caption{The Implicit Profiling Algorithm}
    \label{ag:generalization IP}
    \begin{algorithmic}[1]
    \STATE Initialize $\bt^{(0)}$;
    \STATE Solve $\bl^{(0)}$ from the equation $\bPhi(\bt^{(0)}, \bl^{(0)}) = \mathbf{0}$;
    \REPEAT
        \STATE Update $\bl$ from
        \begin{equation}
            \bl^{(k+1)} = \bl^{(k)} - \left(\frac{\partial \bPhi(\bt^{(k)}, \bl^{(k)})}{\partial \bl}\right)^{-1}\bPhi(\bt^{(k)}, \bl^{(k)});\nonumber
        \end{equation}
        \STATE Solve the implicit gradient $\mathbf{d}^{(k+1)} = \partial\bl^{(k+1)}(\bt^{(k)})/\partial\bt$ from
        \begin{equation}
            \frac{d\bPhi(\bt^{(k)},\bl^{(k+1)})}{d\bt} = \frac{\partial\bPhi(\bt^{(k)}, \bl^{(k+1)})}{\partial \bt} + \frac{\partial \bPhi(\bt^{(k)}, \bl^{(k+1)})}{\partial\bl}\mathbf{d}^{(k+1)} = \mathbf{0}\nonumber
        \end{equation}
        \STATE Compute the implicit profiling Hessian:
        \begin{equation}
        \displayindent0pt
        \displaywidth\textwidth
        \mathbb{H}^{(k+1)}  =   \frac{\partial \bPsi(\bt^{(k)}, \bl^{(k+1)})}{\partial \bt} + \frac{\partial \bPsi(\bt^{(k)}, \bl^{(k+1)})}{\partial \bl}\mathbf{d}^{(k+1)}.
        \nonumber
        \end{equation}
        \STATE Update $\bt$ from
        \begin{equation}
            \bt^{(k+1)} = \bt^{(k)} - \mathbb{H}^{-1}\bPsi(\bt^{(k)}, \bl^{(k+1)});\nonumber
        \end{equation}
    \UNTIL{Convergence}
    \end{algorithmic}
\end{algorithm}
For the initialization of $\bl$, it is recommended to solve the equation $\bPhi(\bt^{(0)}, \bl^{(0)}) = \mathbf{0}$, which can help to improve the convergence speed. However, this calculation may also require high computational cost. In the case that the computational cost is not acceptable, one can also randomly choose an initial value. In each updating iteration, we first update $\bl$, which is denoted by  $\bl^{(k+1)}$. Then, we calculate the implicit profiling Hessian matrix of $\bt$ using the newly updated $\bl^{(k+1)}$, and then get an updated value $\bt^{(k+1)}$. Repeat the iteration steps until convergence, which leads to the final estimates of $\bt$ and $\bl$.

It is notable that, the implicit profiling method accounts for the interaction between $\bt$ and $\bl$ by treating $\bl$ as a function of $\bt$. Consequently, the resulting estimator of $\bt$ should be equal to the estimate implemented by the entire updating method. We summarize this finding in the following two propositions.
\noindent
\begin{proposition}
Assume the objective function $\mathcal{L}$ is strictly convex.
Convergent point of implicit profiling method and that of Newton-Raphson method are identical.
\end{proposition}
\begin{proposition}
For any local quadratic problem $Q$,
implicit profiling method reaches its minimal within two steps.
\end{proposition}
\noindent
The detailed proof of the two propositions are given in Appendix A.1 and A.2, respectively. Propositions 1 and 2 established that the implicit profiling method shared the theoretical properties of the Newton-Raphson method.
By Proposition 1, implicit profiling only converges at the minimum of the convex loss.
By Proposition 2, the convergence is guaranteed when the Newton-Raphson method converges, and the number of iterations taken before converges is comparable to that of the Newton-Raphson method.
Later in the experiments, we found that the fewer number of iterations is the driving factor for implicit profiling method's advantage in run time compared to other iterative methods.
In many cases, single iteration of the implicit profiling can be faster than that of the Newton-Raphson method
when the dependence structure of $\bt$, $\bl$ and the loss function enables the implicit profiling to simplify
the whole Hessian matrix calculation and global value searching of the Newton-Raphson method.
Together with the control on the number of iterations, the implicit profiling method is computationally more efficient than the Newton-Raphson method.

\csection{A TOY EXAMPLE}
\csubsection{Model Description}

To further illustrate the idea of the implicit profiling method, we consider a toy example in this section. Specifically, assume the objective function is $\mL(x,y) = x^2 + y^2 + \alpha xy$, where $-2 < \alpha < 2$. It is noteworthy that, when $\alpha=0$, $x$ and $y$ can be clearly separated from each other. However, when $\alpha \neq 0$, $x$ and $y$ are bundled together in the objective function. Given the objective function is convex, optimization of this objective function needs to solve the following equations:
\begin{equation}
    \label{eq:derivate equation set}
    \left\{
    \begin{aligned}
         & \bPsi(x, y) = \frac{\partial \mL(x, y)}{\partial x} = 2x + \alpha y = 0  \\
         & \bPhi(x, y) = \frac{\partial \mL(x, y)}{\partial y} = 2y + \alpha x = 0.\nonumber
    \end{aligned}
    \right.
\end{equation}

The first-order derivatives $\bPsi(x, y)$ and $\bPhi(x, y)$ control the updating directions for $x$ and $y$, respectively. It is notable that, when $\alpha = 0$, the first-order derivatives only involves one parameter each ($x$ or $y$), which makes the updating directions for $x$ and $y$ independent to each other. However, when $\alpha \neq 0$, the updating direction for one parameter is influenced by the other parameter. In addition, the value of $\alpha$ controls connection between $x$ and $y$. As the absolute value of $\alpha$ becomes larger, the connection between $x$ and $y$ becomes stronger. Consequently, the influence evoked by the other parameter on the updating direction also becomes larger.

To optimize the objective function $\mL(x,y)$, we consider three methods: (1) the Newton-Raphson method, as the representative of the entire updating methods, (2) the naive iteration method, as the representative of recursive updating methods, and (3) our proposed implicit profiling method. Below, we give the updating formulas for each method in details.

\textbf{The Newton-Raphson method.} To apply the Newton-Raphson method for optimization, we need to compute the gradient and Hessian matrix of the objective function with respect to $(x,y)$. Denote $\mathbf{G}(x, y) = (\bPsi(x, y), \bPhi(x, y))^{\top}$ as the gradient. Then, the Hessian matrix can be computed as
\begin{equation}
    \mathbb{H} = \frac{\partial\mathbf{G}(x, y)}{\partial (x,y)}
    =\left(\begin{array}{cc}
        \frac{\partial\bPsi(x,y)}{\partial x} &  \frac{\partial\bPsi(x,y)}{\partial y}\\
        \frac{\partial\bPhi(x,y)}{\partial x} &  \frac{\partial\bPhi(x,y)}{\partial y}
    \end{array}\right)
    =
    \left(\begin{array}{cc}
        2 & \alpha \\
        \alpha & 2
    \end{array}\right).\nonumber
\end{equation}
\noindent
Based on the gradient and Hessian matrix, we can compute the updating formulas for $x$ and $y$ in the Newton-Raphson method, which are present as follows
\begin{equation}
    \label{eq:Newton2}
    \left\{
    \begin{aligned}
        x^{(k+1)} = x^{(k)} - \left(\frac{2}{4-\alpha^2}\bPsi(x^{(k)}, y^{(k)}) - \frac{\alpha}{4 - \alpha^2}\bPhi(x^{(k)}, y^{(k)})\right)  \\
        y^{(k+1)} = y^{(k)} - \left(\frac{\alpha}{4 - \alpha^2}\bPsi(x^{(k)}, y^{(k)}) - \frac{2}{4 - \alpha^2}\bPhi(x^{(k)}, y^{(k)})\right).
    \end{aligned}
    \right.
\end{equation}

\textbf{The naive iteration method.} This method breaks the connection of $x$ and $y$ in the Hessian matrix. In other words, the global Hessian matrix $\mathbb{H}$ is not required. We only use the second-order derivatives of $\bPsi(x,y)$ and $\bPhi(x,y)$ to determine the updating direction for $x$ and $y$, separately. Specifically, the updating formula of the simple iteration method is summarized as
\begin{equation}
    \label{eq:IT2}
    \left\{
    \begin{aligned}
         & y^{(k+1)} = y^{(k)} - \frac{\partial \bPhi(x, y)}{\partial y}(2y^{(k)} + \alpha x^{(k)})  =  y^{(k)} - 2(2y^{(k)} + \alpha x^{(k)})  \\
         & x^{(k+1)} = x^{(k)} - \frac{\partial \bPsi(x, y)}{\partial x}(2x^{(k)} + \alpha y^{(k+1)}) =x^{(k)} - 2(2x^{(k)} + \alpha y^{(k+1)}) . \\
    \end{aligned}
    \right.
\end{equation}
\noindent
We then compare \eqref{eq:IT2} with \eqref{eq:Newton2} in the Newton-Raphson method. In the case that $\alpha=0$, the updating formulas used in the two methods become the same. However, when $x$ and $y$ are bundled together (i.e., $\alpha\neq 0$), the updating formula \eqref{eq:IT2} ignores some information, which should lead to sub-optimal updating directions.

\textbf{The implicit profiling method.} Suppose $x$ is the parameter of interest and $y$ is the nuisance parameter. In the proposed implicit profiling method, we only focus on the efficient estimation of $x$. We first regard $y$ as a function of $x$, which can be found by solving $\partial \bPhi(x,y)/\partial x = 2\partial y/\partial x + \alpha = 0$. Treating $y$ as the function of $x$, which we denote by $y(x)$, the first-order derivative of $\bPsi(x,y)$ can be rewritten as $\partial \bPsi(x,y)/\partial x = 2 + \alpha \partial y/\partial x$. Then, the updating formula in the implicit profiling method can be derived as follows
\begin{equation}
    \label{eq:ip}
    \left\{
    \begin{aligned}
         & y^{(k+1)} = y^{(k)} - \frac{1}{2}(2y^{(k)} + \alpha x^{(k)})             \\
         & x^{(k+1)} = x^{(k)} - \frac{2}{4-\alpha^2}(2x^{(k)} + \alpha y^{(k+1)}).
    \end{aligned}
    \right.
\end{equation}
In each iteration of the implicit profiling method, we would first obtain $y^{(k+1)}$ and then plug-in the corresponding value into the updating formula for $x^{(k+1)}$. By simple calculation, the updating formula for $x^{(k+1)}$ with $y^{(k+1)}$ plugged in can be derived as
\begin{equation}
    x^{(k+1)} = x^{(k)} - (\frac{2}{4 - \alpha^2}\bPsi(x^{(k)}, y^{(k)}) - \frac{\alpha}{4 - \alpha^2}\bPhi(x^{(k)}, y^{(k)})),\nonumber
\end{equation}
which exactly the same as the updating formula of $x$ in \eqref{eq:Newton2}. This finding suggests that, the implicit profiling method and the Newton-Raphson method share the same updating direction for $x$, which verifies our statement in Proposition 1.

\csubsection{Simulation Studies}

To evaluate the estimation performance of the proposed implicit profiling method, we conduct simulation studies under the objective $\mL(x, y) = x^2 + y^2 + \alpha xy$. Specifically, we consider different settings of $\alpha$ to control the connection between $x$ and $y$. For illustration purpose, we let $\alpha$ vary from 0 to 1.8, with a step of 0.2. Under each value of $\alpha$, we optimize the objective function using the Newton-Raphson method, the naive iteration method and the implicit profiling method. In consideration of randomness, we choose 100 different initial values under each estimation method. Specifically, the initial values $x^{(0)}$ and $y^{(0)}$ are calculated using the following formula:
\begin{equation}
\begin{split}
         & x^{(0)} = \frac{1}{\sqrt{(2 - \frac{\alpha^2}{2})}}(\sqrt{C(1-\frac{\alpha}{2})}\cos{\gamma} + \sqrt{C(1+\frac{\alpha}{2})}\sin{\gamma})  \\
         & y^{(0)} = \frac{1}{\sqrt{(2 - \frac{\alpha^2}{2})}}(\sqrt{C(1-\frac{\alpha}{2})}\cos{\gamma} - \sqrt{C(1+\frac{\alpha}{2})}\sin{\gamma}),\nonumber
\end{split}
\end{equation}
where $C = k^2$ with $k = 1, 2, ..., 10$ and $\gamma = 2\pi\times\{0.1,0.2,...,1\}$.

In this simple toy example, all three methods result in the same optimal values of $x$. Therefore, we only focus on the computational efficiency of different methods, which is measured by the average number of iterative steps consumed by each method. We evaluate the computational efficiency from two perspectives: (1) the influence of $\alpha$, which characterizes the connection between $x$ and $y$; and (2) the influence of $C$, which depicts the distance between the initial value and the optimal value.

\begin{figure}[H]
    \centering
    \subfigure[Average steps with different $\alpha$]{
    \centering
    \includegraphics[width = 0.48\textwidth]{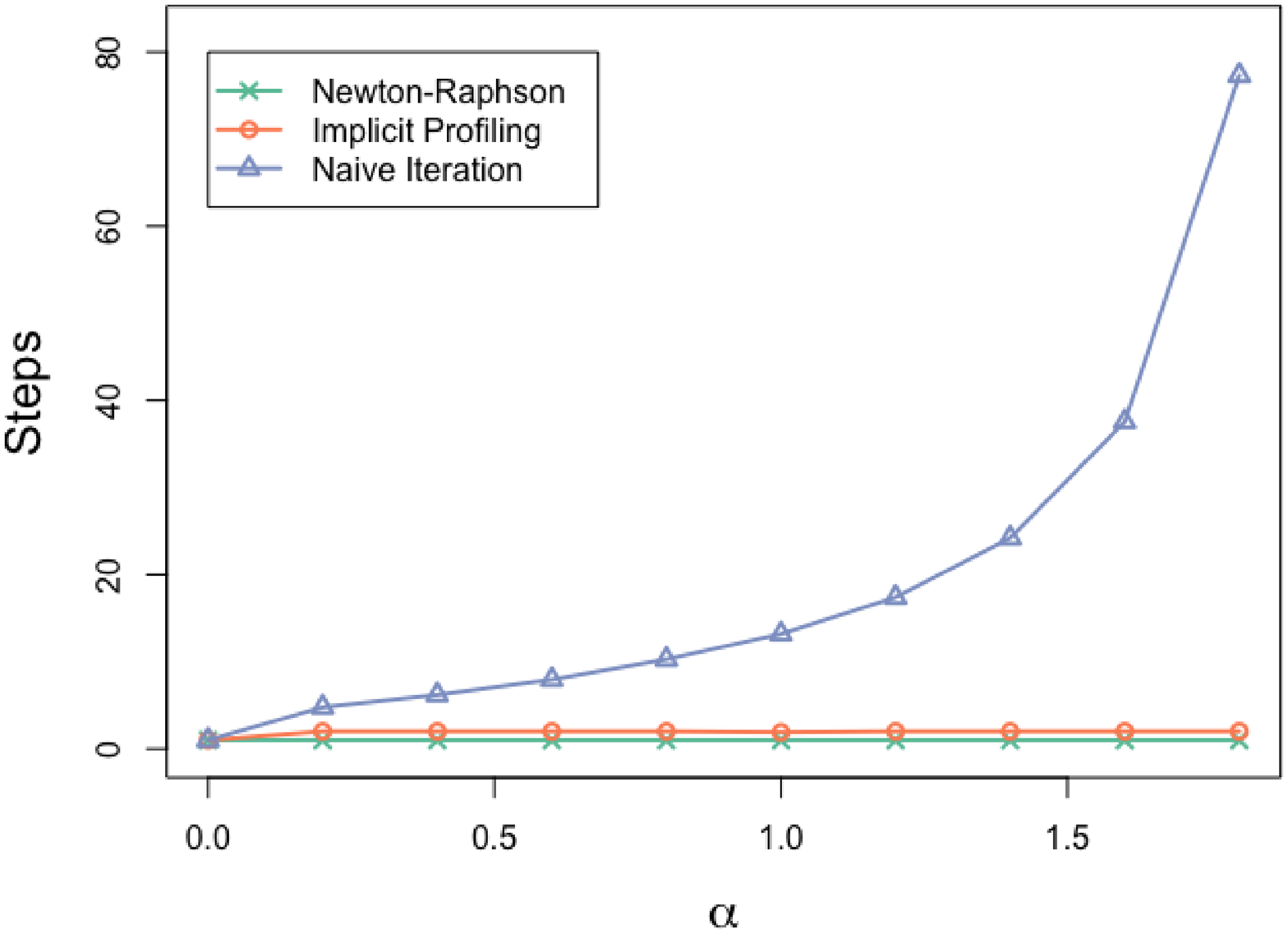}
    }
    \subfigure[Average steps with different $C$]{
    \centering
    \includegraphics[width = 0.48\textwidth]{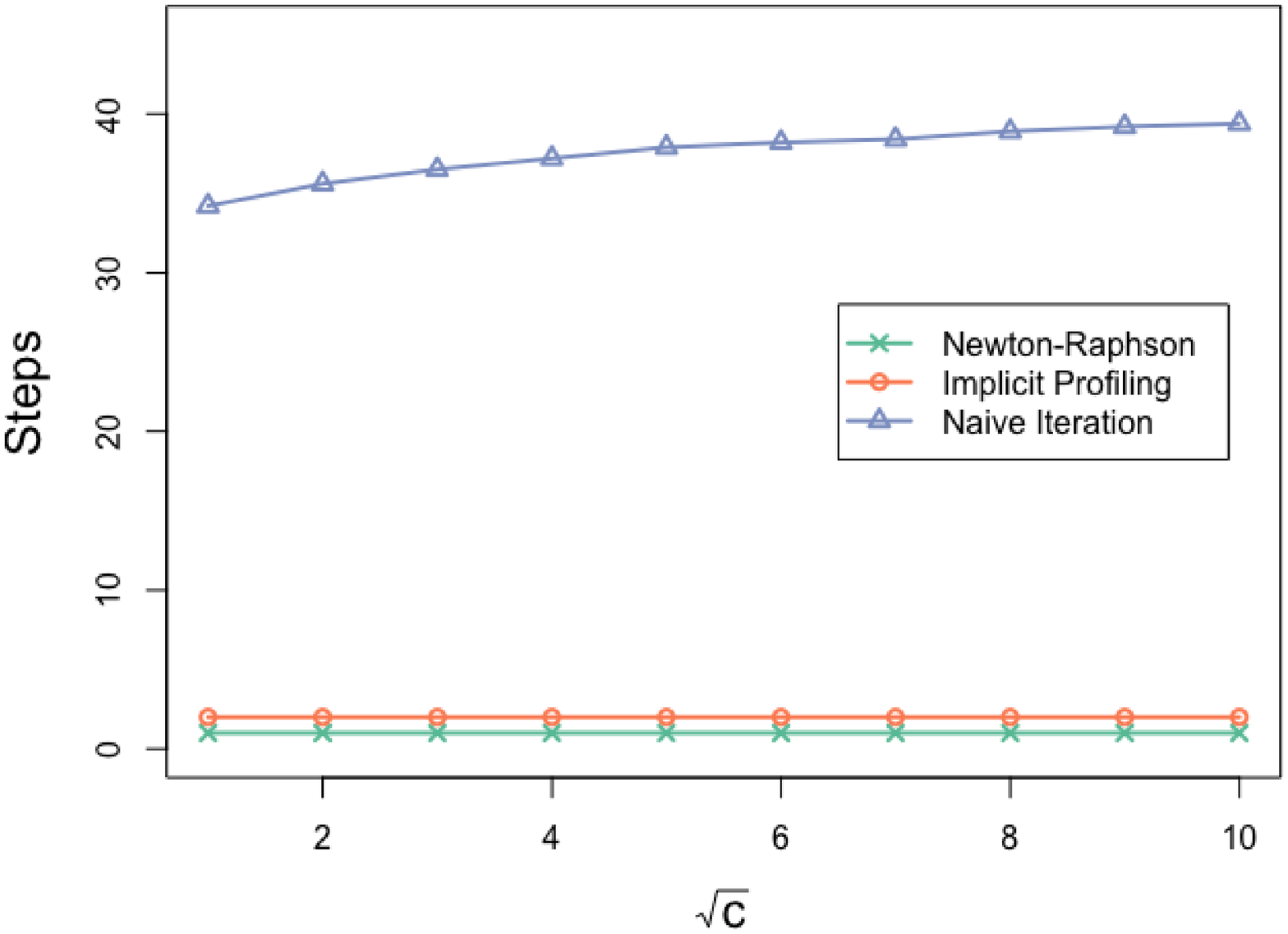}}
    \caption{The average number of iterative steps consumed by the Newton-Raphson method, the naive iteration method and the implicit profiling method. The subfigure (a) explores the influence of different values of $\alpha$, while the  subfigure (b) explores the influence of different values of $C$ when fix $\alpha=1.6$.}
    \label{fig:Steps}

\end{figure}

We first focus on the influence of $\alpha$. The average number of iterative steps consumed by the three methods under each value of $\alpha$ are present in Figure \ref{fig:Steps}(a). As shown, the Newton-Raphson method only takes one step to reach convergence; while the implicit profiling method takes two steps. This is because, the two methods share the same updating formula for $x$. However, the implicit profiling method requires another additional step to update $y$. Compared with the two methods, the naive iteration method has the worst computational efficiency. When $\alpha>0$, the average number of iterative steps consumed by the naive iteration method is larger than the implicit profiling method and the Newton-Raphson method. In addition, it requires more steps to converge when $\alpha$ becomes larger. This finding suggests that, then the connection of $x$ and $y$ becomes stronger, the naive iteration method losses more information in its updating formula and consequently it behaves less computationally efficient.

Then we focus on the influence of initial value on the computational efficiency of different methods. We fix $\alpha=1.6$ and vary $C = k^2$ with $k = 1, 2, ..., 10$. As the increase of $C$, the initial values of $x_0$ and $y_0$ have a bigger distance from the optimal values. Under each setup of $C$, we calculate the number of iterative steps consumed by each method, which is shown in Figure \ref{fig:Steps}(b). As shown, under different initial values, the Newton-Raphson method always converges by just one step, and the implicit profiling method can reach its convergence with two steps. Neither of the two methods is influenced by the initial values. Compared with the fast convergence of the Newton-Raphson method and the implicit profiling method, the naive iteration method consumes a large number of steps to convergence. In addition, as the distance between the initial values and optimal values gets larger, the naive iteration method requires more steps to converge.
\begin{figure}[H]
    \centering
    \subfigure[Naive Iteration]{
        \label{subfigure:IT}
        \includegraphics[width=0.45\textwidth]{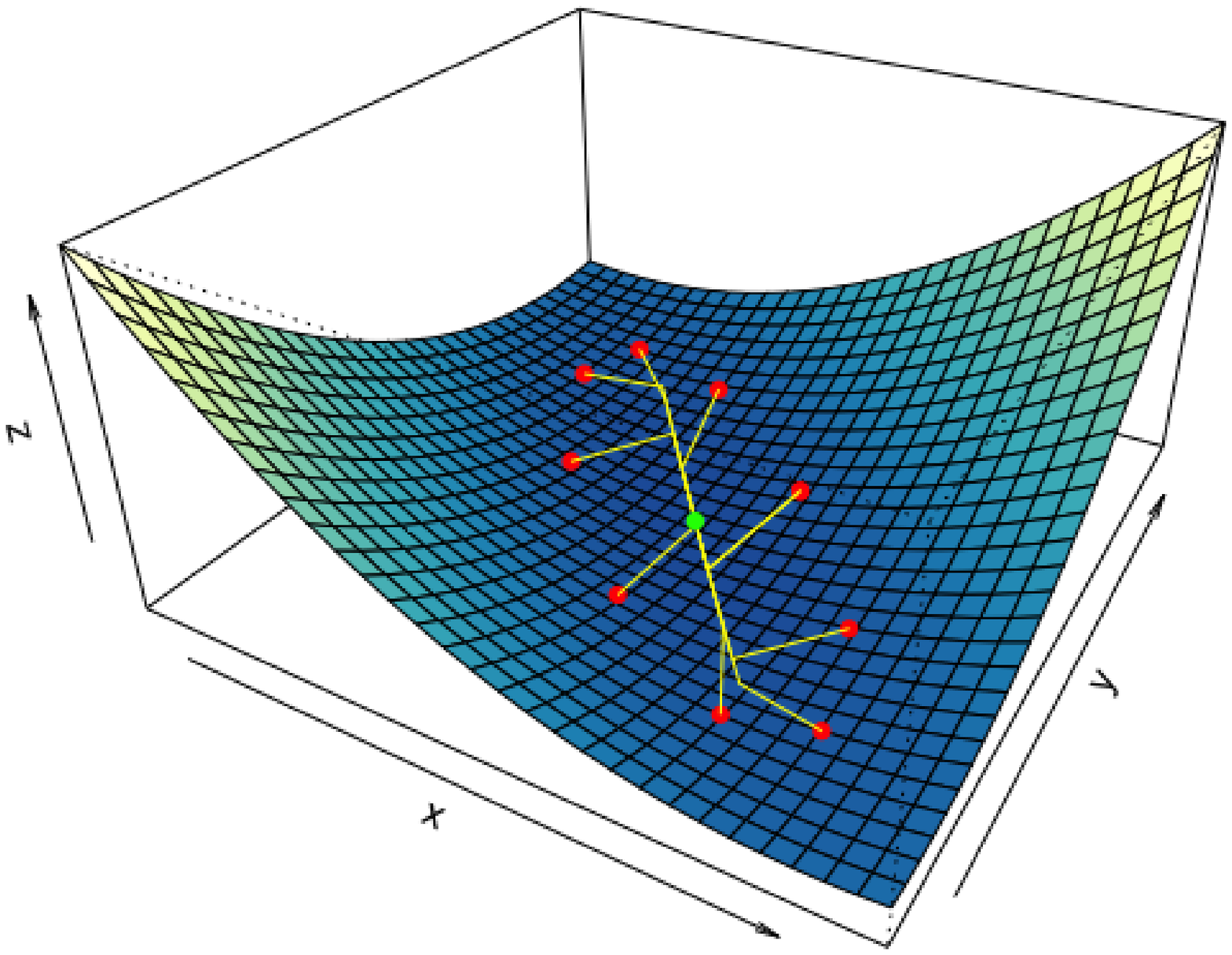}}
    \subfigure[Implicit Profiling]{
        \label{subfigure:IP}
        \includegraphics[width=0.45\textwidth]{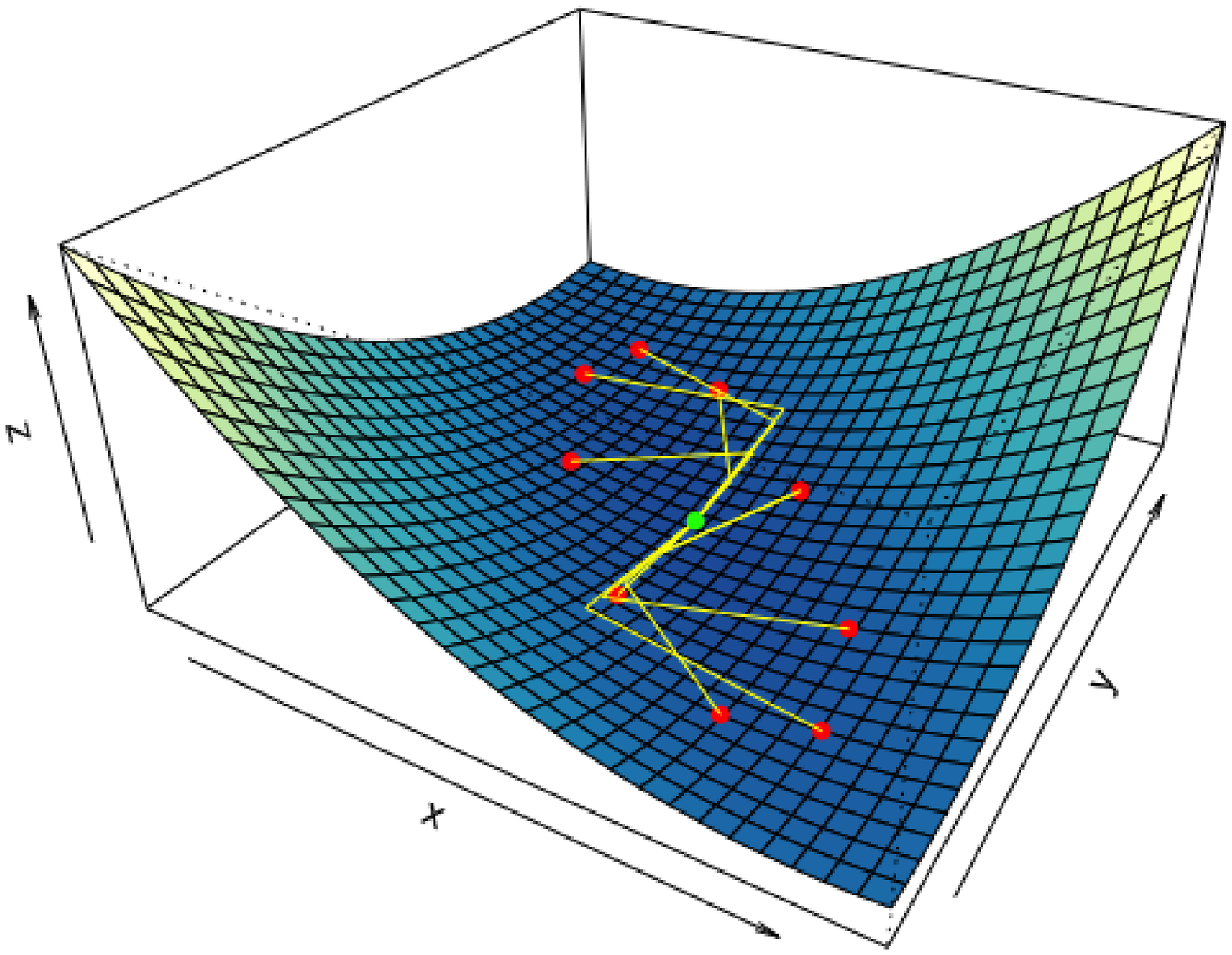}}
    \caption{The convergence paths of the naive iteration method and the implicit profiling method under $\alpha = 1.6$ and $C = 4$.}
    \label{figure:path}
\end{figure}

To further illustrate the computational efficiency of the implicit profiling method over the naive iteration method, we compare their convergence paths under $\alpha = 1.6$ and $C = 4$. Figure \ref{figure:path} presents the convergence paths of the two methods, where the green points denote the true values and the red points denote the initial values. As shown in Figure \ref{figure:path}, the implicit profiling method only takes two steps to converge. However, the naive iteration method takes about 30 iterative steps to converge.
Compared with the implicit profiling method, the naive iteration method dose find the correct updating direction after its first update. However, the updating step size used in the naive iteration method is not efficient, which makes it suffer from a larger number of iteration steps to converge. This is mainly due to the lost information it ignores when evaluating the second-order derivatives.

\csection{APPLICATION: SEMI-PARAMETRIC TRANSFORMATION MODEL}
\csubsection{Model Description}

In this section, we focus on a semiparametric transformation model \citep{1998Generalized,chanRiskPredictionImperfect2021} to illustrate the application of implicit profiling method. Specifically, for the $i$th subject with $1 \leq i \leq n$, let $T_i$ and $C_i$ denote the true event time and follow up time, respectively. Let $\delta_i = I(T_i \leq C_i)$ define whether the event has occurred by the end of the follow up time, where $I(\cdot)$ is an indicator function. To help estimate the occurrence risk, define $\mathbf{Z}_i$ as a $p$-dimensional covariate vector associated with subject $i$. Then, the objective function to be estimated is
\begin{equation}
    \label{eq:Risk prediction model}
    E(\delta_i \vert C_i, \mathbf{Z}_i) = P(T_i \leq C_i \vert C_i, \mathbf{Z}_i) = \pi(\bl(t) + \bt^\top \mathbf{Z}_i).\nonumber
\end{equation}
Here, $\bt$ is the $p$-dimensional parameter vector, which is the research interest. $\bl(\cdot)$ is an unspecified smooth increasing function, which is the nonparametric component with infinite dimension. $\pi(\cdot) = \exp(\cdot)/(1 + \exp(\cdot))$ denotes the logit function. To estimate the smooth function $\bl(\cdot)$, the kernel estimation method is applied \citep{chanRiskPredictionImperfect2021}. Therefore, the objection function with respect to $\bt$ and $\bl$ are derived as follows:
\begin{equation}
    \label{eq:transformation}
    \left\{
    \begin{aligned}
& \bPsi(\bt, \bl, C_i) = n^{-1}\sum_{j=1}^n K_h(C_j - C_i)[\delta_j - \pi\{\bl(C_i)+\bt^\top Z_j\}] = 0                                                                                       \\
     & \bPhi(\bt, \bl) = n^{-1} \sum_{j=1}^n Z_j[\delta_j - \pi\{\bl(C_j)+\bt^\top Z_j\}]  = 0,
    \end{aligned}
    \right.
\end{equation}
where $K_h(\cdot)$ is the kernel function with the bandwidth $h$. Further denote $\bPsi(\bt, \bl) = (\bPsi(\bt, \bl, C_1), \bPsi(\bt, \bl, C_2), ..., \bPsi(\bt, \bl, C_n))^\top$. It is notable that, $\bl$ is the smooth function, whose dimension is the same as the sample size $n$. Then, the first-order derivative $\bPsi(\bt,\bl)$ is also a $n$-dimensional vector. $\bt$ and $\bPhi(\bt, \bl)$ are both $p$-dimensional vectors.

\csubsection{Application of the Implicit Profiling Method}

In the semiparametric transformation model, the nonparametric component $\bl(t)$ is a smooth function, whose dimension should grow with the number of sample size. Therefore, using the Newton-Raphson method to estimate the model would be clumsy, because the computation of the inverse of Hessian matrix is very computationally expensive. To address this issue, we can apply the implicit profiling method to increase the computational efficiency, as it can guarantee $\bt$ to convergence as the same direction as the Newton-Raphson method.

We follow \eqref{eq:ip_update2} to implement the implicit profiling method on the semiparametric transformation model. Based on \eqref{eq:transformation}, we first derive the second-order derivatives of the objective function as follows
\begin{equation}
\label{eq:tmp}
    \left\{
    \begin{aligned}
        &\frac{d\bPsi(\bt, \bl, C_i)}{d\bt} =  - \sum_{i=1}^n Z_i\pi'\{\bl(C_i)+\bt^\top Z_i\}
  \{Z_i - \nabla_{\bt} \bl(C_i)\}^\top\\
  &\frac{d\bPhi(\bt, \bl)}{d\bl} = -\sum_{j=1}^n K_h(C_j - C_i)\pi'\{\bl(C_i)+\bt^\top Z_j\}
\{\nabla_{\bt} \bl(C_i) + Z_j\}.
    \end{aligned}
    \right.
\end{equation}
The key of the implicit profiling method is to regard $\bl(\cdot)$ as the function of $\bt$, and then profile out the nonparametric component. To this end, define $\bl(C_i)$ as the function of $\bt$. Then we can calculate the first-order derivative of $\bl(C_i)$ with respect to $\bt$ by solving $\partial \bPhi(\bt, \bl)/\partial\bl = \mathbf{0}$. Define $\mathbf{d}_i = d \bl(C_i)/d\bt$ as the corresponding derivative used in the $k$th iteration. We can compute $\mathbf{d}_i$ as follows
\begin{equation}
    \mathbf{d}_i^{(k)}  = -\frac{\sum_{j=1}^n K_h(C_j - C_i)\pi'\{\bl_i^{(k+1)}+\bt^{(k)\top} Z_j\} \nonumber
        Z_j}{\sum_{j=1}^n K_h(C_j - C_i)\pi'\{\bl_i^{(k+1)}+\bt^{(k)\top} Z_j\}},
\end{equation}
where $\bl_i$ is the $i$th component of $\bl$. Substitute $\mathbf{d}_i^{(k)}$ into $d\bPsi(\bt, \bl, C_i)/d\bt$, and we can get the implicit profiling Hessian matrix of $\bt$
\begin{equation}
    \mathbb{H}^{(k)}  =  \sum_{i=1}^n Z_i Z_i^\top\pi'\{\bl_i^{(k+1)}+\bt^{(k)\top}Z_i\} + \sum_{i=1}^{n}Z_i\mathbf{d}_i^{(k)\top}\pi'\{\bl_i^{(k+1)}+\bt^{(k)\top}Z_i\}.\nonumber
\end{equation}
With the Hessian matrix $\mathbb{H}^{(k)}$ and the formula $\partial \bPhi(\bt, \bl)/\partial\bl$ derived in \eqref{eq:tmp}, we summarize the updating equations of the implicit profiling method, which iterates $\bt$ and $\bl$ separately.
\begin{equation}
    \left\{
    \begin{aligned}
        \bl^{(k+1)} & = \bl^{(k)} - \left[\sum_{j=1}^n K_h(C_j - C_i)\pi'\{\bl_i^{(k)}+\bt^{(k)\top} Z_j\}\right]^{-1}\bPhi(\bt^{(k)}, \bl^{(k)})                                                                                                                   \\
        \bt^{(k+1)} & = \bt^{(k)} -\mathbb{H}^{(k)-1}\bPsi(\bt^{(k)}, \bl^{(k+1)}).\nonumber
    \end{aligned}
    \right.
\end{equation}

\csubsection{Simulation Studies}

We conduct some simulation experiments to evaluate the performance of the implicit profiling method on the semiparametric transformation model. Assume the whole sample size $n=(500,1000)$. Further assume the number of covariates $p=10$, and the corresponding parameter $\bt = (\theta_j)_{j=1}^{p}= (0.7,0.7,0.7,-0.5,-0.5,-0.5,0.3,0.3,0.3,0)^{\top}$. We then generate the data $(\delta_i, C_i, Z_i)$ with $1 \leq i \leq n$ in the following procedure. For the $i$th subject, each of its covariates $\mathbf{Z}_i$ is generated from a standard normal distribution. Then we generate $u_i$ from $\mathbf{U}(0, 1)$ and $C_i$ from $\mathbf{U}(0, 12)$. Based on $u_i$, we compute $T_i  = 4\exp((\ln(u_i)-\ln(1-u_i) - \mathbf{Z}_i^{\top}\bt)/3)$. Finally, the binary indicator $\delta_i$ is computed as $\delta_i = I(T_i \leq C_i)$.

We repeat the data generation process for $B=100$ times. For each generated dataset, we apply the Newton-Raphson method, the naive iteration method, as well as the implicit profiling method for estimation. The Newton-Raphson method and the naive iteration method are implemented using the R package \emph{nleqslv}, which is particularly designed for solving a system of nonlinear equations. The three methods share the same initial values and tolerance criterion for convergence. For each method, we define $\hat{\bm{\theta}}^{(b)}=(\hat{\theta}_{j}^{(b)})_{j=1}^{p}$ as the estimator for $\bm{\theta}$ in the $b$th replication ($1\leq b \leq B$). Then, to evaluate the estimation efficiency of each estimator, we define the MSE for $\hat{\theta}_{j}$, namely, $\text{MSE}(\hat{\theta}_{j})=B^{-1}\sum_{b=1}^{B}(\hat{\theta}_{j}^{(b)}-\theta_j)^2$. Finally, we compute the averaged MSE of all $\theta_j$s as the final performance measure, namely, $\text{MSE}=p^{-1}\sum_{j=1}^{p}\text{MSE}(\hat{\theta}_{j})$. Except for the MSE, we also evaluate the computational efficiency of the three methods, which are measured by the number of iterations and the total computational time.

Table \ref{tab:result} shows the simulation results for three methods in details, where the MSE values, the averaged computational time and averaged iteration steps are reported. From the results in Table \ref{tab:result}, we can draw the following conclusions. First, the three methods have achieved the same estimates for $\bt$, given they have obtained the same MSE values in the experiments. These results are in accordance with those in Section 3, which indicates that the three methods can research the same optimal value and thus the corresponding estimates have the same statistical efficiency. In addition, with the increase of sample size $n$, the MSE values of all three methods become smaller. Second, among all the methods, the implicit profiling method has the smallest computational times, implying the computational efficiency of this method. As for the naive iteration method, it needs much more iteration rounds to converge. This is because it has lost information in computing the second-order derivatives. Compared with the implicit profiling method and the naive iteration method, the Newton-Raphson method needs the least iteration steps. However, because the Newton-Raphson method needs to calculate the whole Hessian matrix, it has the highest computational time.
\begin{table}[H]
    \centering
    \setstretch{1}
    \caption{The simulation results under the semiparametric transformation model. The MSE values, averaged computational time (in seconds) and averaged iteration rounds for the Newton-Raphson method, the naive iteration method and the implicit profiling method are reported.}
    \label{tab:result}
    \begin{tabular}{ccccc}
        \toprule
        N                      & Method & RMSE    & Time    & Iterations  \\
        \midrule
        \multirow{3}{*}{500}   &Implicit Profiling     & 0.462 & 0.32 & 9.17 \\
                               &Naive Iteration     & 0.462 & 2.37 & 18.54 \\
                               &Newton-Raphson & 0.462 & 76.24  & 6.94\\
        \midrule
        \multirow{3}{*}{1,000} & Implicit Profiling     & 0.324 & 1.15  & 9.28 \\
                               & Naive Iteration     & 0.324 & 15.02 & 17.98 \\
                               &Newton-Raphson & 0.324 &565.56&7.23\\
        \bottomrule
    \end{tabular}
\end{table}


\csection{APPLICATION: SEMIPARAMETRIC GARCH-M MODEL}


\csubsection{Model Description}

In this section, we apply the implicit profiling method on the semiparametric GARCH-in-mean model, which we refer to as GARCH-M for short \citep{Christensen2012}. This model has been popularly used in financial time series analysis. The key of the GARCH-M model is to introduce the variance of the time series into the mean function. Specifically, denote $y_t$ as a time series variable and $\sigma_t^2$ is the variance of $y_t$. The GARCH-M model has the following form:
\begin{equation}
    \label{eq:GARCH-M}
    \left\{
    \begin{aligned}
         & y_t = \bl(\sigma_t^2) + \epsilon_t                             \\
         & \sigma_t^2 = \omega + \alpha y_{t-1}^2 + \beta \sigma_{t-1}^2,
    \end{aligned}
    \right.
\end{equation}
where $\bl(\cdot)$ is some smooth function, $\epsilon_t$ is the random noise, and $\bt = (\omega, \alpha, \beta)^{\top}$ is the parameter of interest. In the GARCH-M model, the time series variable $y_t$ is influenced by its variance $\sigma_t^2$ through $\bl(\cdot)$, and the variance $\sigma_t^2$ is in turn influenced by the past information $y_{t-1}$ and $\sigma_{t-1}^2$. Based on the GARCH-M model shown in \eqref{eq:GARCH-M}, the quasi-maximum likelihood function can be constructed as follows
\begin{equation}
\label{eq:quasi-maximum}
    f(\bt, \bl(\sigma_t^2)) = -\frac{1}{2} \sum_{t=1}^T \ln (\sigma_t^2) - \frac{1}{2} \sum_{t=1}^T\frac{(y_t-\bl(\sigma_t^2))^2}{\sigma_t^2},
\end{equation}
where $T$ denotes the time span. To estimate the GARCH-M model, one needs to maximize the above quasi-maximum likelihood function, which is the objective function in the GARCH-M model.

In the previous literature, two methods are often applied to solve this objective function, i.e., the backfitting method \citep{Christensen2012} and the SP-MBP \citep{Frazier2018A}. However, neither of the two methods exploit the data information sufficiently. In other words, both the backfitting method and SP-MBP method are approximation solutions. To address this issue, we apply the implicit profiling method to estimate the GARCH-M model. Specifically, the smooth function $\bl(\cdot)$ is estimated by B-spline. Then, the implicit profiling method focuses on solving the following equations
\begin{equation}
    \left\{
    \begin{aligned}
         & \bPsi(\bt, \hat\bl) = \Psi_{1, T}(\bt, \hat{\bl}(\hat\sigma_t))-\Psi_{2, T}(\bt, \hat{\bl}(\hat\sigma_t)) = 0 \\
         & \bPhi(\bt, \hat\bl) = \hat\bl(\sigma_t^2) - \sum_{i=1}^{R_T} \gamma_i B_{i, 2}(\sigma_t^2) = 0,\nonumber
    \end{aligned}
    \right.
\end{equation}
where $\Dot{\bl}(\cdot)$ is the first-order derivative of $\bl(\cdot)$ with respect to $\bt$, $B_{i,2}(\sigma_t^2)$ is B-spline of order 2 with $t_i\leq\sigma_t^2\leq t_{i+1}$, $\Psi_{1, T}(\bt, \hat{\bl}(\hat\sigma_t))$ and $\Psi_{2, T}(\bt, \hat{\bl}(\hat\sigma_t))$ are defined as follows
\begin{equation}
    \begin{split}
        &\Psi_{1, T}(\bt, \hat{\bl}(\hat\sigma_t)) = \frac{1}{2} \sum_{t=1}^T \left( \frac{1}{\hat{\sigma}_t^2(\bt)} - \frac{\Tilde{\epsilon}_t^2(\bt)}{(\hat{\sigma}_t^2(\bt))^2}\right)\frac{\partial \hat{\sigma}_t^2(\bt)}{\partial \bt} \\
        &\Psi_{2, T}(\bt, \hat{\bl}(\hat\sigma_t)) = \sum_{t = 1}^T\frac{\Tilde{\epsilon}_t(\bt)}{\hat{\sigma}_t^2(\bt)}\Dot{\Tilde{\bl}}(\hat{\sigma}_t^2(\bt)),\nonumber
    \end{split}
\end{equation}
where $\hat{\epsilon}_t(\bt) = y_t - \bl(\hat{\sigma}_t^2(\bt), \bt)$ and $\hat{\sigma}_t^2(\bt) = \omega + \alpha y_{t-1}^2 + \beta \hat{\sigma}_{t-1}^2(\bt)$. Together with $\Psi_{1, T}(\bt, \hat{\bl}(\hat\sigma_t))$ and $\Psi_{2, T}(\bt, \hat{\bl}(\hat\sigma_t))$, $\bPsi(\bt, \hat\bl)$ is the derivative of $f(\bt, \bl(\sigma_t^2(\bt)))$ with respect to $\bt$. To get the B-spline approximation $\hat\bl(\cdot)$, let $[a, b]$ be the range of $\sigma_t^2 = \sigma_t^2(\bt)$. Let the knots $\{t_1, t_2, ..., t_{R_T}\}$ partition $[a, b]$, where the total number of knots satisfies $R_T = O(T^\nu)$, $\max_{1 \leq j \leq R_T} t_j = O(T^{-\nu})$ and $\nu \in (0, 1/2)$. The first-order derivative $\hat{\bl}(\sigma_t^2)$ is then obtained by estimating the corresponding coefficients $\gamma_i (i = 0,...,R_T)$, which only requires least-squares calculations.

In the GARCH-M model, we can get the explicit expression of $\hat\bl$, which is a function of $\bt$. Then $\hat\bl$ can be updated directly in the implicit profiling method. The entire updating formula used in the implicit profiling method is shown below:
\begin{equation}
    \left\{
    \begin{aligned}
         & \hat\bl^{(k+1)}(\hat\sigma_t^2) =  \sum_{i=1}^{R_N} \gamma_i B_{i, 2}(\hat\sigma_t^2(\bt^{(k)}))                                                                                                                                                   \\
         & \bt^{(k+1)} = \bt^{(k)} -\left( \frac{\partial \bPsi(\bt^{(k)}, \hat\bl^{(k+1)})}{\partial \bt} + \frac{\partial \bPsi(\bt^{(k)}, \hat\bl^{(k+1)})}{\partial \hat\bl}\frac{d\bl^{(k+1)}}{d\bt}\right)^{-1}\bPsi(\bt^{(k)}, \hat\bl^{(k+1)}).\nonumber
    \end{aligned}
    \right.
\end{equation}

\csubsection{Simulation Studies}

We present a variety of simulation studies to evaluate the performance of the implicit profiling method on the GARCH-M model. Following \cite{Frazier2018A}, we consider two data generating processes as follows
\begin{equation}
    \begin{split}
        &A: y_t = \sigma_t^2 + 0.5\sin (10\sigma_t^2) + \sigma\epsilon_t, \sigma_t^2 = \omega + \alpha y_{t-1}^2 + \beta\sigma_{t-1}^2 \\
        &B: y_t = 0.5\sigma_t^2 + 0.1\sin (0.5 + 20\sigma_t^2) + \sigma\epsilon_t, \sigma_t^2 = \omega + \alpha y_{t-1}^2 + \beta\sigma_{t-1}^2,\nonumber
    \end{split}
\end{equation}
where $\epsilon_t$ is the white noise generated from the standard normal distribution. We fix the parameters $\omega=0.01$ and $\alpha=0.1$, but set $\beta=0.68$ in the setup A and $\beta=0.80$ in the setup B. In each setup, we consider the time span $T=(500,1,000)$. The data generation process is also repeated by $B=100$ times.

For each generated dataset, we apply the implicit profiling method for estimation. For comparison purpose, we also estimate the GARCH-M model using the backfitting method (BF) and the SP-MBP method. All methods share the same initial values and tolerance criterion for convergence. In all the methods, $\bl(\cdot)$ is approximated by B-spline and the number of knots $R_T$ is set as $T^{3/20}$. For one particular method (i.e., IP, BF and SP-MBP), we define $\hat{\bm{\theta}}^{(b)}=(\hat{\theta}_{j}^{(b)})_{j=1}^{p}$ as the estimator in the $b$-th ($1\leq b \leq B$) replication. Then, to evaluate the estimation efficiency of each estimator, we calculate the bias as $\flat=\bm{\theta}-\bar{\bm{\theta}}$, where $\bar{\bm{\theta}}=B^{-1}\sum_b\hat{\bm{\theta}}^{(b)}$. Then, we compute the Monte Carlo standard deviation of $\hat\beta^{(b)}$, which is calculated by $\mbox{SE}=\{B^{-1}\sum_b (\hat{\bm{\theta}}^{(b)}-\bar{\bm{\theta}})^2\}^{1/2}$. In addition, we compute the mean absolute error (MAE), $B^{-1}\sum_b \vert \hat\bt^{(b)} - \bt\vert$, and root mean squared error (RMSE), $\{B^{-1}\sum_b(\hat\bt^{(b)} - \bt)^2\}^{1/2}$, for evaluation, while the root mean squared error is the same as Section 4.

Table \ref{tab:my_label} presents the simulation results under experimental setup A, and the corresponding results under experimental setup B are present in Appendix A.3 to save space. In general, the simulation results in the two experimental setups have similar patterns. It is obvious that, the implicit profiling method yields more precise estimation results by achieving smaller MAE and RMSE in most cases. In particular, $\beta$ is an important parameter in the GARCH-M model, as it describes the lag effect of variance. Compared with the backfitting method and the SP-MBP method, our proposed implicit profiling method can achieve more accurate estimation results on this parameter.
\begin{table}[h]
\setstretch{1}
    \caption{The simulation results under the GARCH-M model with experimental setup A. The bias, standard deviation, MAE and RMSE for IP, backfitting and SP-MBP methods are reported. }
    \label{tab:my_label}
    \resizebox{\textwidth}{!}{
        \begin{tabular}{llrrrlrrr}
            \toprule
            Method                 & N=500 & ${\omega}$ & ${\alpha}$ & ${\beta}$ & N = 1,000 & ${\omega}$ & ${\alpha}$ & ${\beta}$ \\
            \midrule
            \multirow{4}{*}{IP}    & BIAS  & 0.0019     & -0.0005    & -0.0119   & BIAS      &  0.0012     & 0.0002    & -0.0070   \\
                                   & SE   & 0.0083     & 0.0196     & 0.0528    & SE       & 0.0067     & 0.0145     & 0.0450    \\
                                   & MAE   & 0.0063     & 0.0154     & 0.0415    & MAE       & 0.0046     & 0.0112     & 0.0319    \\
                                   & RMSE  & 0.0085     & 0.0196     & 0.0541    & RMSE      & 0.0068    & 0.0145     & 0.0455   \\
            \midrule
                                    & N=500 & ${\omega}$ & ${\alpha}$ & ${\beta}$ & N = 1,000 & ${\omega}$ & ${\alpha}$ & ${\beta}$ \\
            \midrule
            \multirow{4}{*}{BF}    & BIAS  & 0.0016     & -0.0000    & -0.0088   & BIAS      & 0.0014     & 0.0004    & -0.0065   \\
                                   & SE   & 0.0092     & 0.0232     & 0.0770    & SE       & 0.0067     & 0.0170     & 0.0556    \\
                                   & MAE   & 0.0067     & 0.0185     & 0.0589   & MAE       & 0.0049     & 0.0134     & 0.0436   \\
                                   & RMSE  & 0.0093     & 0.0232     & 0.0775    & RMSE      & 0.0068     & 0.0170     & 0.0560    \\
            \midrule
                                   & N=500 & ${\omega}$ & ${\alpha}$ & ${\beta}$ & N = 1,000 & ${\omega}$ & ${\alpha}$ & ${\beta}$ \\
            \midrule
            \multirow{4}{*}{SP-MBP} & BIAS  & 0.0023     & 0.0017     & -0.0157   & BIAS      & 0.0012     & 0.0016     & -0.0079   \\
                                   & SE   & 0.0106     & 0.0246     & 0.0863    & SE       & 0.0066     & 0.0180     & 0.0583    \\
                                   & MAE   & 0.0073     & 0.0196     & 0.0652    & MAE       & 0.0048     & 0.0141     & 0.0457   \\
                                   & RMSE  & 0.0108    & 0.0247     & 0.0876   & RMSE      & 0.0067     & 0.0180     & 0.0588    \\
            \bottomrule
        \end{tabular}}
\end{table}

Next, we focus on the computational efficiency of these methods. Table \ref{tab:my_label3} presents the averaged computational time (in seconds) and the number of total iterations consumed by each method in different experimental settings. As shown, the SP-MBP method is the most computationally expensive by consuming the most computational time. It is followed by the implicit profiling method. Among the three methods, the backfitting method is the most computationally efficient. The computational advantage of the backfitting method mainly results from its simple structure and the least information used to update. However, when focusing on the iteration steps, Table \ref{tab:my_label3} shows that the implicit profiling method converges more quickly than the other two methods by consuming the smallest number of total iterations.
\begin{table}[h]
    \centering
    \setstretch{1}
    \caption{The averaged computational time (in seconds) and the number of total iterations consumed by backfitting (BF), SP-MBP and implicit profiling methods.}
    \label{tab:my_label3}
    \begin{tabular}{lrrrr}
        \toprule
                 & (500, A) & (1000, A) & (500, B) & (1000, B) \\
        \midrule
        &\multicolumn{4}{c}{Panel A: Computational Time}\\
        \cline{2-5}
        IP       & 0.3956   & 0.4243   & 0.4928   & 0.5460    \\
        BF       & 0.1654   & 0.2049    & 0.1157   & 0.1675    \\
        SPMBP    & 1.7836   & 1.5726    & 1.0643   & 1.1731    \\
        \midrule
        &\multicolumn{4}{c}{Panel B: Total Iterations}\\
        \cline{2-5}
        IP       & 47.10  & 39.54    & 56.94  & 50.56    \\
        BF       & 163.37   & 112.35    & 105.13   & 87.55    \\
        SPMBP    & 196.02   & 129.06    & 107.69   & 92.98    \\
        \bottomrule
    \end{tabular}
\end{table}

In summary, the above simulation results indicate that, the implicit profiling method can obtain more accurate estimation results of $\bm{\theta}$ than backfitting and BP-MBP methods. Although the implicit profiling method has higher computational time than the backfitting method, it shows great advantages against the BP-MBP method in computational efficiency. Note that the total computational time for the GARCH-M model is relatively small and dose not increase dramatically with the time span. Therefore, the estimation accuracy should be the main focus in this application. From this perspective, the implicit profiling method has shown a good performance.

\csection{CONCLUSION AND DISCUSSION}

We propose in this work an implicit profiling method for estimation of semiparametric models with bundled parameters. We classify semiparametric models with bundled parameters into the explicitly bundled type and implicitly bundled type, both of which can be solved by using the implicit profiling method. The the new gradient and Hessian matrix of the parametric component can be computed by plugging in the nonparametric function. Using this way, the relationship between the parametric component and nonparametric component is taken into account. We show theoretically that, the implicit profiling method can converge to the optimal point under a convex objective function. This property guarantees the statistically efficiency of the implicit profiling method. It also behaves computationally efficient when compared with previous entire updating methods and recursive methods. We first illustrate the computational advantages of implicit profiling method by a toy example. Then we take the semiparametric transformation model as the implicitly bundled example. Compared with the Newton-Raphson method and naive iteration method, our proposed implicit profiling method shows great computational advantages by consuming the least computational time. Finally, the semiparametric GARCH-M model is considered as an example of explicit bundled type. Compared with two state-of-the-art methods, the implicit profiling method achieves the highest statistical efficiency as well as comparable computational speed.

Finally, we conclude this work by discussing the relationship between the implicit profiling method with the Expectation-Maximization (EM) algorithm. The EM algorithm is a widely used method in practice. It recursively processes the E-step and M-step to make estimation. However, this method often takes a long time to converge, especially in complicated situations with censoring and missing values. In addition, it can suffer from the curse of dimensionality. Our proposed implicit profiling method can be regarded as a substitute to the EM algorithm. In the application of the implicit profiling method, the E-step can be regarded as the nonparametric component while the M-step can be regarded as the parametric component. Consequently, the implicit profiling method can accelerate the computational efficiency in solving the EM designed problems.

\noindent
\textbf{Code Variability}

We implement the proposed implicit profiling method for semiparametric models in an R package called \emph{SemiEstimate}, which can be downloaded directly from CRAN (The Comprehensive R Archive Network). It contains all codes for the simulation examples, which are easy for users to reproduce. For user friendly design, the numerical derivative approximate is applied. Then only the initial values and the target equation functions are required to implement our method via the function "semislv()". The Jacobin matrix is also allowed to provide. An advanced DIY mode for researches is also supported in the package, which allows users to omit repeated calculation to save time. More source codes can be found in the GitHub repository "JinhuaSu/SemiEstimate".

\noindent
\textbf{Acknowledgement}

The work is supported by National Natural Science Foundation of China (72001205, 11971504), fund for building world-class universities (disciplines) of Renmin University of China, the Fundamental Research Funds for the Central Universities and the Research Funds of Renmin University of China (2021030047), Foundation from Ministry of Education of China (20JZD023), Ministry of Education Focus on Humanities and Social Science Research Base (Major Research Plan 17JJD910001).

\newpage
\bc
{\bf\large APPENDIX}
\ec
\renewcommand{\theequation}{A.\arabic{equation}}
\setcounter{equation}{0}

\scsection{Appendix A.1: Proof of Proposition 1}\label{appendixa1}
Below, we will show that implicit profiling has the same convergence point as Newton-Raphson method. Since we assume that $\mathcal{L}$ is strictly convex, the hessian matrix of $\mathcal{L}$ has eigenvalues strictly bounded away from zero.
We can easily prove that there will only exists a stationary point for the Newton-Raphson method. For proof purpose, we will first prove that the convergence point of Newton-Raphson method solves the implicit profiling problem, and then prove that the solution of implicit profiling is a stationary point of Newton-Raphson method.

\noindent\textbf{Statement 1: Convergence point of Newton-Raphson solves implicit profiling problem}

It's known that the convergence of Newton-Raphson method to the stationary point is guaranteed. And for implicit profiling updating formulas \eqref{eq:ip_update2}, the stationary point $(\bt^\star, \bl^\star)$, where $\bPsi(\bt^\star, \bl^\star) = 0$ and $\bPhi(\bt^\star, \bl^\star) = 0$, satisfies
\begin{equation}
    \left\{
    \begin{aligned}
          &\bl^\star = \bl^\star - \left(\frac{\partial \bPhi(\bt^\star, \bl^\star)}{\partial \bl}\right)^{-1}\bPhi(\bt^\star, \bl^\star)\\
          &\bt^\star = \bt^\star -\left( \frac{\partial \bPsi(\bt^\star, \bl^\star)}{\partial \bt} + \frac{\partial \bPsi(\bt^\star, \bl^\star)}{\partial \bl}\frac{\partial\bl^\star}{\partial\bt}\right)^{-1}\bPsi(\bt^\star, \bl^\star).
    \end{aligned}\nonumber
    \right.
\end{equation}

Thus, when iteration reaches stationary point, the iteration will stop, and implicit profiling converges.

\noindent\textbf{Statement 2: Solution of implicit profiling is a stationary point of Newton-Raphson}

Now, we denote the stationary point of implicit profiling method as $(\bt^\star, \bl^\star)$.
By \eqref{eq:ip_update2}, $(\bt^\star, \bl^\star)$ satsifies
\begin{equation}
    \left\{
    \begin{aligned}
          &\bl^\star =\bl^\star - \left(\frac{\partial \bPhi(\bt^\star, \bl^\star)}{\partial \bl}\right)^{-1}\bPhi(\bt^\star, \bl^\star)\\
          &\bt^\star = \bt^\star -\left( \frac{\partial \bPsi(\bt^\star, \bl^\star)}{\partial \bt} + \frac{\partial \bPsi(\bt^\star, \bl^\star)}{\partial \bl}\frac{\partial\bl^\star}{\partial\bt}\right)^{-1}\bPsi(\bt^\star, \bl^\star).
    \end{aligned}\nonumber
    \right.
\end{equation}
Suppose the dimensions for the two parts of the parameters are
$\bt \in \mathbb{R}^p$ and $\bl \in \mathbb{R}^q$.
Denote the unit sphere in $\mathbb{R}^d$ as $\mathbb{S}^{d-1}$
Since the objective function is strictly convex, the diagonal block in its Hessian $\partial \bPhi(\bt^\star, \bl^\star)/\partial \bl$ is thus positive definite,
\begin{align*}
    \inf_{\mathbf{v} \in \mathbb{S}^{q-1}} \mathbf{v}^\top \frac{\partial \bPhi(\bt^\star, \bl^\star)}{\partial \bl} \mathbf{v}
    = & \inf_{\mathbf{v} \in \mathbb{S}^{q-1}} (\mathbf{0}^\top_p, \mathbf{v}^\top)
    \left(
    \begin{array}{cc}
        \frac{\partial^2}{\partial \bt\bt^\top} \mathcal{L}(\bt^\star, \bl^\star) &
        \frac{\partial^2}{\partial \bt\bl^\top} \mathcal{L}(\bt^\star, \bl^\star) \\
        \frac{\partial^2}{\partial \bl\bt^\top} \mathcal{L}(\bt^\star, \bl^\star)&
        \frac{\partial^2}{\partial \bl\bl^\top} \mathcal{L}(\bt^\star, \bl^\star)
    \end{array}
    \right) \left(
    \begin{array}{c}
       \mathbf{0}_p\\
       \mathbf{v}
    \end{array}
    \right) \\
 \ge & \inf_{\mathbf{u} \in \mathbb{S}^{p+q-1}}
\mathbf{u}^\top \left(
    \begin{array}{cc}
        \frac{\partial^2}{\partial \bt\bt^\top} \mathcal{L}(\bt^\star, \bl^\star) &
        \frac{\partial^2}{\partial \bt\bl^\top} \mathcal{L}(\bt^\star, \bl^\star) \\
        \frac{\partial^2}{\partial \bl\bt^\top} \mathcal{L}(\bt^\star, \bl^\star)&
        \frac{\partial^2}{\partial \bl\bl^\top} \mathcal{L}(\bt^\star, \bl^\star)
    \end{array}
    \right)\mathbf{u}.
\end{align*}
Thus, we deduce from the stationary point equations that
$$
\bPhi(\bt^\star, \bl^\star) = \mathbf{0}_{q}.
$$
Similar to $\partial \bPhi(\bt^\star, \bl^\star)/\partial \bl$,
another diagonal block in the Hessian of $\mathcal{L}$,
$\partial \bPsi(\bt^\star, \bl^\star)/\partial \bt$,
is also positive definite.
The matrix
$$
\frac{\partial \bPsi(\bt^\star, \bl^\star)}{\partial \bl}\frac{\partial\bl^\star}{\partial\bt}
= \frac{\partial^2}{\partial \bt\bl^\top} \mathcal{L}(\bt^\star, \bl^\star) \left\{\frac{\partial^2}{\partial \bl\bl^\top} \mathcal{L}(\bt^\star, \bl^\star)\right\}^{-1}
\frac{\partial^2}{\partial \bl\bt^\top} \mathcal{L}(\bt^\star, \bl^\star)
$$
is symmetric, thus positive semi-definite.
As the sum of a positive definite matrix and a positive semi-definite matrix,
$$
\frac{\partial}{\partial \bt}\bPsi(\bt^\star, \bl^\star)
+ \frac{\partial \bPsi(\bt^\star, \bl^\star)}{\partial \bl}\frac{\partial\bl^\star}{\partial\bt}
$$
is also positive definite.
Then, we deduce from the stationary point equations that
$$
\bPsi(\bt^\star, \bl^\star) = \mathbf{0}_p.
$$
As mentioned above, for convex function with positive definite hessian matrix, there only exits a stationary point for Newton-Raphson method solving
the first order condition,
$$
\bPhi(\bt^\star, \bl^\star) = \mathbf{0}_{q}, \;
\bPsi(\bt^\star, \bl^\star) = \mathbf{0}_p.
$$
So the solution of implicit profiling converges is a stationary point, which is also the convergence of Newton-Raphson method.

\scsection{Appendix A.2: Proof of Proposition 2}\label{appendixa2}

Below, we show the computational efficiency of the implicit profiling method. For illustration purpose, we prove the computational efficiency of the method in quadratic case, where it only takes two steps to converge. We start with the behaviour of the Newton-Raphson method. Recall $\bm\beta = (\bt^\top, \bl^\top)^\top$ and the updating formula in Newton-Raphson method is $\betat{k+1} = \betat{k}-(\partial \bm{G(\betat{k})}/\partial \bm\beta)^{-1}\bm{G}(\betat{k})$. Assume the dimensions of $\bt$ and $\bl$ are $p$ and $q$, respectively. Then $\bm\beta$ is a $p+q$-dimensional vector. Consider a generic quadratic function $Q$ mapping $\mathbb{R}^{p+q}$ to $\mathbb{R}^1$, i.e.,
\begin{equation}
    Q(\bm\beta) = g^\top\bm\beta + \frac{1}{2}\bm\beta^\top \bm{H}\bm\beta,\nonumber
\end{equation}
where $\bm{H}$ is assumed to be nonsingular and positive semi-definite. Then the gradient of $Q(\bm\beta)$ can be computed as $\bm{G}(\bm\beta) = g + \bm H\bm\beta$, and the Hessian matrix of $Q(\bm\beta)$ is $\mathbb{H} = \bm H$. Consequently, the updating formula for Newton-Raphson method is
\begin{equation}
\label{eq:A1}
    \betat{k+1} = \betat{k} - \mathbb{H}^{-1}\bm G(\betat{k})
    = \betat{k} - \bm H^{-1}(g + \bm H\betat{k}).
\end{equation}

Assume there exits a stationary point, i.e., $\bm\beta^\star = (\bt^{\star\top}, \bl^{\star\top})^\top$ and $\bm G(\bm\beta^\star) = \bm 0$. We show next that, for any $\betat{0}$ in the neighborhood of $\bm\beta^\star$, the Newton-Raphson method only takes one step to converge to $\bm\beta^\star$. Specifically, by \eqref{eq:A1}, we have $\betat{1} = \betat{0} - \bm H^{-1}(g + \bm H\betat{0})$. Then the gradient at the point $\betat{1}$ becomes
\begin{equation}
\bm G(\betat{1}) = \bm G(\betat{0} - \bm H^{-1}(g + \bm H\betat{0}))= g + \bm H(\betat{0} - \bm H^{-1}(g + \bm H\betat{0})) = \bm 0.\nonumber
\end{equation}
Therefore, we know $\betat{1}$ is the stationary point.

Next, we focus on the implicit profiling method. Before giving the computational details of the implicit profiling method, we rewrite the above equations again using $\bt$ and $\bl$. Specifically, we can rewrite the quadratic function $Q$ as follows
\begin{equation}
    Q(\bt, \bl) = \left(
    \begin{array}{c}
        g_1\\
         g_2
    \end{array}\right)^\top\left(\begin{array}{c}
        \bt\\
         \bl
    \end{array}\right)
    + \frac{1}{2}\left(
    \begin{array}{c}
        \bt\\
        \bl
    \end{array}\right)^\top\left(\begin{array}{cc}
        \bm H_{11} &\bm H_{12} \\
        \bm H_{21} &\bm H_{22}
    \end{array}\right)
    \left(
    \begin{array}{c}
        \bt \\
        \bl
    \end{array}\right),\nonumber
\end{equation}
where $g=(g_1,g_2)^{\top}$ and $\bm H=[\bm H_{11},\bm H_{12};\bm H_{21},\bm H_{22}]$. Then the gradient $\bm{G}(\bt, \bl)$ is rewritten as
\begin{equation}
    \bm G(\bt, \bl) = \left(\begin{array}{c}
        \bm G_1(\bt, \bl)  \\
        \bm G_2(\bt, \bl)
    \end{array}\right) = \left(
    \begin{array}{c}
        g_1\\
         g_2
    \end{array}\right) + \left(\begin{array}{cc}
       \bm H_{11} & \bm H_{12} \\
       \bm H_{21} &\bm H_{22}
    \end{array}\right)\left(
    \begin{array}{c}
        \bt \\
        \bl
    \end{array}\right),\nonumber
\end{equation}
and the updating formula in \eqref{eq:A1} is rewritten as follows,
\begin{equation}
\begin{aligned}
\left(
    \begin{array}{c}
        \btt{k+1}  \\
        \blt{k+1}
    \end{array}\right) &= \left(
    \begin{array}{c}
        \btt{k}  \\
        \blt{k}
    \end{array}\right) - \left(\begin{array}{cc}
       \bm H_{11} & \bm H_{12} \\
       \bm H_{21} &\bm H_{22}
    \end{array}\right)^{-1}\left(\begin{array}{c}
        \bm G_1(\bt, \bl)  \\
        \bm G_2(\bt, \bl)
    \end{array}\right) \\
    &= \left(
    \begin{array}{c}
        \btt{k}  \\
        \blt{k}
    \end{array}\right) - \left(
    \begin{array}{cc}
    \mathbf{F}     &  -\mathbf{F}\mathbf{H}_{11}\mathbf{H}_{22}^{-1} \\
    -\mathbf{H}_{22}^{-1}\mathbf{H}_{21}\mathbf{F}     & \mathbf{H}_{22}^{-1}(\mathbf{I} + \mathbf{H}_{21}\mathbf{F}\mathbf{H}_{12}\mathbf{H}_{22}^{-1})
    \end{array}
    \right)\left(\begin{array}{c}
        \bm G_1(\bt, \bl)  \\
        \bm G_2(\bt, \bl)
    \end{array}\right), \nonumber
\end{aligned}
\end{equation}
where $\mathbf{F} = (\bm H_{11} - \bm H_{12}\bm H_{22}^{-1}\bm H_{21})^{-1}$. To compute the implicit profiling Hessian matrix, we need first solve $d \bl/d\bt$ from
\begin{equation}
    \frac{\d\bm G_2(\bt, \bl))}{\d \bt} = \bm H_{21} + \bm H_{22}\frac{\d \bl}{\d \bt} = \mathbf{0}.\nonumber
\end{equation}
Then the implicit profiling Hessian matrix can be derived as
\begin{equation}
    \mathbb{H}_{\bt} = \frac{\d \bm G_1(\bt, \bl)}{\d \bt} = \bm H_{11} + \bm H_{12}\frac{\d \bl}{\d \bt} = \bm H_{11} - \bm H_{12}\bm H_{22}^{-1}\bm H_{21} = \mathbf{F}^{-1}.\nonumber
\end{equation}
By substituting $\mathbb{H}_{\bt}$ into \eqref{eq:ip_update2}, we can derive the updating formula for the implicit profiling method
\begin{equation}
\label{eq:lambdaA}
    \left\{
    \begin{aligned}
    &\blt{k+1} = \blt{k} - \parfrac{\bm G_2(\btt{k},\blt{k})}{\bl}^{-1}\bm G_2(\btt{k}, \blt{k}) = \blt{k} - \bm H_{22}^{-1}\bm G_2(\btt{k}, \blt{k})\\
    &\btt{k+1} = \btt{k} - \mathbb{H}_{\bt}^{-1} \bm G_1(\btt{k},\blt{k+1}) = \btt{k} - \bm F\bm G_1(\btt{k},\blt{k} - \bm H_{22}^{-1}\bm G_2(\btt{k}, \blt{k}) ).
    \end{aligned}
    \right.
\end{equation}
Further substituting $\bm G_1(\btt{k}, \blt{k})$ and $\bm G_2(\btt{k}, \blt{k})$ into the updating formula $\btt{k+1}$ and we can obtain:
\begin{equation}
\begin{aligned}
\label{eq:IPA}
     \btt{k+1} &= \btt{k} - \bm F (g_1 + \bm H_{11}\btt{k} + \bm H_{12}\blt{k}) + F\bm H_{12}\bm H_{22}^{-1}(g2 + \bm H_{22}\blt{k} + \bm H_{21}\btt{k})\\
     &= \btt{k} - (\bm F \bm G_1(\btt{k}, \blt{k}) - \bm F \bm H_{12}\bm H_{22}^{-1}\bm G_2(\btt{k}, \blt{k})).\nonumber
\end{aligned}
\end{equation}
It is notable that, \eqref{eq:IPA} is the same as the updating formula used in the Newton-Raphson method. Therefore, for any initial value $\btt{0}$ in the neighborhood of $\bt^\star$, it will converge to $\bt^\star$ at the first step as similar as the Newton-Raphson method. However in the implicit profiling method, we need an additional step to update $\bl$. Specifically, define $\blt{1}$ as the current value in the first step. By the updating formula in \eqref{eq:lambdaA}, we have $\blt{2} = \blt{1} - \bm H_{22}^{-1}\bm G_2(\bt^\star, \blt{1})$. Finally, we substitute $(\bt^\star, \blt{2})$ into the gradient of $\bm G_2$ to show $\blt{2}$ is also the stationary point.
\begin{equation}
\begin{aligned}
      \bm G_2(\bt^{\star}, \blt{2}) &= \bm G_2(\bt^{\star}, \blt{1} - \bm H_{22}^{-1}\bm G_2(\bt^{\star}, \blt{1})) \\
     &= g_2 + \bm H_{21}\bt^{\star} + \bm H_{22}(\blt{1} - \bm H_{22}^{-1}\bm G_2(\bt^{\star}, \blt{1})) \\
     &= g_2 + \bm H_{21}\bt^\star + \bm H_{22}\blt{1} - \bm G_2(\bt^\star, \blt{1})
     = \mathbf{0}.\nonumber
\end{aligned}
\end{equation}
This completes the proof of Proposition 2.

\scsection{Appendix A.3: Simulation Results for the GARCH-M Model under Experimental Setup B}\label{appendixa3}
\begin{table}[H]
\setstretch{1}
    \caption{The simulation results under the GARCH-M model with experimental setup B. The bias, standard deviation, MAE and RMSE for IP, backfitting and SP-MBP methods are reported. }
    \label{tab:my_label2}
    \resizebox{\textwidth}{!}{
        \begin{tabular}{llrrrlrrr}
            \toprule
            Method                 & N=500 & ${\omega}$ & ${\alpha}$ & ${\beta}$ & N = 1,000 & ${\omega}$ & ${\alpha}$ & ${\beta}$ \\
            \midrule
            \multirow{4}{*}{IP}    & BIAS  & 0.0034     & -0.0056    & -0.0297   & BIAS      & 0.0014     & -0.0033    & -0.0108   \\
                                   & SE   & 0.0105     & 0.0369     & 0.1115    & SE       & 0.0057     & 0.0260     & 0.0676    \\
                                   & MAE   & 0.0064     & 0.0302     & 0.0751   & MAE       & 0.0039     & 0.0207     & 0.0490    \\
                                   & RMSE  & 0.0110     & 0.0373    & 0.1153   & RMSE      & 0.0059     & 0.0262     & 0.0684    \\
             \midrule
                                    & N=500 & ${\omega}$ & ${\alpha}$ & ${\beta}$ & N = 1,000 & ${\omega}$ & ${\alpha}$ & ${\beta}$ \\
            \midrule
            \multirow{4}{*}{BF}    & BIAS  & 0.0030     & -0.0016    & -0.0286   & BIAS      & 0.0016     & -0.0013   & -0.0142   \\
                                   & SE   & 0.0092     & 0.0366    & 0.1062    & SE       & 0.0059     & 0.0264     &0.0716    \\
                                   & MAE   & 0.0059     & 0.0295     & 0.0726    & MAE       & 0.0041     & 0.0214     & 0.0512    \\
                                   & RMSE  & 0.0097     & 0.0366     & 0.1100    & RMSE      & 0.0061     & 0.0265     & 0.0730    \\
            \midrule
                                   & N=500 & ${\omega}$ & ${\alpha}$ & ${\beta}$ & N = 1,000 & ${\omega}$ & ${\alpha}$ & ${\beta}$ \\
            \midrule
            \multirow{4}{*}{SP-MBP} & BIAS  & 0.0031     & 0.0011    & -0.0297   & BIAS      & 0.0016     & -0.0009    & -0.0150   \\
                                   & SE   & 0.0093    & 0.0371     & 0.1080    & SE       & 0.0061     & 0.0268     & 0.0735   \\
                                   & MAE   & 0.0060     & 0.0298     & 0.0740    & MAE       & 0.0041     & 0.0217     & 0.0523   \\
                                   & RMSE  & 0.0098     & 0.0371     & 0.1119   & RMSE      & 0.0063     & 0.0268     & 0.0750    \\
            \bottomrule
        \end{tabular}}
\end{table}

\bibliographystyle{asa}
\bibliography{reference}
\end{document}